\magnification=1200 \hsize=6.2 truein \vsize=8.9 truein
\def\ni{\noindent}

\def\m{{\cal M}_4}\def\z{Z\!\!\!Z}\def\l{\Lambda}
$\,$
\hfill SWAT/2001/286

\ni {\bf Spin and Electromagnetic Duality: An Outline}
\bigskip
\ni Talk presented at the 15th Anniversary Meeting of 
Dirac Medallists,
at Abdus Salam ICTP, Trieste, November 2000.
\bigskip
\ni{\bf David I Olive},
\medskip
\ni Physics Department,
 University of Wales Swansea, Swansea SA2 8PP
\bigskip
\ni {\bf Introduction}
\bigskip

Quantum electromagnetic duality is a new sort of quantum symmetry principle
that manifests itself in  quantum field theories of the unified type.
Certain physical quantities exhibit invariance or
 covariance under the action of
an infinite discrete group, such as the modular group, acting on 
a dimensionless coupling constant. It seems to me to be important to clarify
what is going on by studying various examples of this effect.

Rather than talk about the usual scenarios,
 supersymmetric gauge theories [1] or string theories,
 I shall talk about a different and relatively simple setup:
 a fixed four-dimensional background  space-time that may be
complicated topologically. Nevertheless it is assumed to be closed,
 compact, smooth and oriented. It supports non-singular Maxwell field
 strengths that may be dynamical and, in any case,
 have to allow the presence of complex quantum wave
functions for charged particles. These wave functions may be either scalar
or spinor, according to the spin of the corresponding 
 charged particle, and the
consequences will be distinguished and compared.

My discussion falls into two parts following the
two papers written in collaboration
with Marcos Alvarez [2], [3]. The first part deals with the Dirac quantisation condition
for magnetic fluxes, that is, the consistency condition for the
 existence of the complex wave functions. There is an anomalous effect
when complex spinor wave functions are considered which may
lead to half-integral rather than integral
 fluxes through certain sorts of two-cycle. In four dimensional space-times
when the fine structure constant is dimensionless and 
traditional electromagnetic duality holds,
 there is a simple characterisation
of these anomalous two-cycles, namely that they are precisely the ones with
 odd self-intersection number.
These results are already known to pure mathematicians but we shall
find a more physical language
for explaining them and talk about a slight generalisation,
 what we have called a \lq\lq  quantum Stokes' theorem'', that holds
when the Maxwell gauge potential or connection can only be defined locally,
in patches, because of the complication of the background space-time
 topology.

In the second part, this effect is shown to tie in nicely
 with quantum electromagnetic duality.
 A sort of generalised partition function for the Maxwell theory
can be evaluated using semiclassical methods, following E Witten [4]
 and E Verlinde [5]. The stationary points are labelled by the values of the quantised
fluxes and so the sum over them yields a result proportional to a generalised
 theta function. This transforms nicely under the action of a subgroup of the modular group
acting on the dimensionless coupling constant of the theory.

When the space-time background is such that the electromagnetic fields
can support scalar and spinor wave functions simultaneously, the range of the sum is an
even integral lattice. There is only one partition function and
it transforms nicely under the full modular group of electromagnetic duality
transformations.

When the background space-time possesses the aforementioned anomalous
two-cycles  there is a disagreement
between the quantum consistency conditions for the fluxes needed 
to allow scalar and spinor
wave functions. Therefore  two distinct partition functions
 have to be considered.
 In one the range of summation
is an odd unimodular lattice and the other this lattice 
displaced by one half of what is known
as the characteristic vector of the odd unimodular lattice.
 By the Atiyah-Singer index theorem
this encodes the self-intersection numbers mentioned above.
Separately these two partition functions 
each transform nicely under a subgroup of the modular group.
A simple and general construction with these lattices shows how the action
 of the full modular group
is recovered by a mixing of these two, together with
 a third that arises naturally. It is this unexpected
 effect that explains the above title,
spin and electromagnetic duality.
\bigskip
\ni {\bf Fluxes and Homology}
\bigskip
Faraday's concept of electromagnetic fluxes through two-dimensional surfaces
led mathematicians to a more precise language that will prove very relevant
and useful physically and will fit in well with the ideas of Dirac
that came later.

Open and closed surfaces are distinguished. An open surface $\Sigma$ has a boundary, denoted $\partial\Sigma$. If this boundary
 vanishes the surface is closed and called a cycle.
Two two-cycles are equivalent, i.e. homologous,
 if they differ by the boundary of
a three-dimensional object.
$$\Sigma\sim \Sigma'\leftrightarrow \Sigma-\Sigma'=\partial\alpha$$
 These equivalence classes form a
homology class and the set of these classes in a given
background four-dimensional space-time, $\m$, is denoted
$H_2(\m,\z)$. The electromagnetic field strength tensor
in space-time defines a closed two-form, denoted $F$, where $dF=0$.
Then the magnetic flux through the two-cycle
 $\Sigma$ can be written as $\int_{\Sigma}F$.
By virtue of Stokes' theorem,
 this is unchanged with respect to alteration of $\Sigma$ to a homologous
$\Sigma'$, as above, and $F$ to a cohomologous $F'$, that is such that
$F'=F+dB$ and so is also closed as the exterior derivative, 
$d$, like the boundary operator,  $\partial$, squares to zero.

Thus the notion of flux possesses a degree of invariance and moreover
can be subjected to an operation of addition in a natural way:
$$\int_{\Sigma_1}F+\int_{\Sigma_2}F=\int_{\Sigma_1+\Sigma_2}F.$$
So the set $H_2(\m,\z)$ forms an abelian group under addition
 that reflects the structure
of the space-time. Although this group usually possesses an infinite number of
elements it is finitely generated. The elements of finite
order  give rise to vanishing Maxwell fluxes and so are irrelevant for the present
physical purposes. Because they form an invariant subgroup they can be divided out to leave a group described by a finite number, $b_2$, of copies of the integers:
$$H_2(\m,\z)/FINITE\equiv\z^{b_2}.$$
Thus the resultant group can be thought of as a lattice of dimension $b_2$.
The positive integer $b_2$ is known as the second Betti number of $\m$.

The other Betti numbers for $\m$, $b_0, b_1,b_3$ and $b_4$ can be
 defined similarly, by adjusting the dimension of the cycles considered.
 One of the consequences of the Poincar\'e duality
property of space-time $\m$ is that the Betti numbers are equal in complementary  pairs,
$b_1=b_3$ and $b_0=b_4$. We shall   assume  space-time is connected so
that  $b_0=b_4=1$ but not that it is simply connected (so $b_1$
need not vanish).
A particular  linear combination is familiar,
 $$\chi(\m)=b_0-b_1+b_2-b_3+b_4=2(1-b_1)+b_2$$
 and is called the Euler number of $\m$. Unlike
  the individual Betti numbers, it
is local in the sense of being expressible as the integral of a local quantity
over $\m$. Hence it is likely to play a special role in local quantum field theory
and this will be confirmed. There is one other such local quantity,
called the Hirzebruch signature, denoted
$$\eta(\m)=b_2^+-b_2^-\qquad\hbox{where}\qquad b_2^++b_2^-=b_2.$$
Definitions of $b_2^{\pm}$ will be given two sections later.
\bigskip
\ni{\bf Dirac Quantisation Condition and the Quantum Stokes Relation}
\bigskip
We now seek electromagnetic field configurations on $\m$ that permit
a consistent definition of a complex scalar wave function, $\phi(x)$, there.
Since this brings into play the gauge potential one-form, $A$, obtained
from $F$ by integrating $F=dA$ we are forced to work
 in topologically trivial neighbourhoods covering $\m$. In overlaps between
a pair of neighbourhoods the two choices of $A$ and $\phi$ should be patched
 together
by the $U(1)$ gauge transformations:
$$A\rightarrow A+d\chi, \quad\hbox{so}\quad F\rightarrow F,$$
$$\phi(x)\rightarrow e^{{iq\chi(x)\over\hbar}}\phi(x)$$
where the gauge function $\chi$ is real and $q$ is the electric charge
of the corresponding particle. Consistency conditions arise in triple 
overlap regions and imply, as shown by Orlando Alvarez [6], the
Dirac quantisation conditions for the magnetic fluxes [7]:
$$q\int_{\Sigma}F\quad\in \quad 2\pi\hbar\z.$$

Actually something more can then be shown; if $\Sigma$ is now open:
$$e^{{iq\over\hbar}\int_{\Sigma}F}=e^{{iq\over\hbar}\int_{\partial\Sigma}A}.$$
The point here is that, although the exponent on the left hand side is
unambiguous, that on the right hand side has a quantised ambiguity arising
from choices made in the patching procedure
that is precisely removed by taking the exponential. Because of the necessary
presence
of Planck's constant we have referred to this as the quantum Stokes' relation.
Of course, if $\Sigma$ is closed the right hand side equals unity and Dirac's
condition is recovered. Notice also that taking the charge $q$ to vanish
yields a trivial identity.

It is worth noting how economical this argument is. There is no need of a
 metric on space-time, nor any special dimension. No equations
of motion are needed and there is no restriction on the topology of $\Sigma$.
If it is simply a two-sphere the argument reduces to the familiar one of Wu
and Yang [8], involving two hemispherical neighbourhoods.

On the other hand the quantum description of
 charged spin $1/2$ particles requires complex spinor wave functions. Then
 the fact that the spinor representation of
the orthogonal or Lorentz group (whichever is appropriate) is two-valued
introduces sign choices in the patching procedure.
This follows through to yield an unexpected sign
 in the quantum Stokes' relation
which now reads:
$$e^{{iq\over\hbar}\int_{\Sigma}F}=(-1)^{w(\Sigma)}e^{{iq\over\hbar}\int_{\partial\Sigma}A}.$$
This extra sign, $(-1)^{w(\Sigma)}$, is intrinsic to $\Sigma$,
 that is independent
of the choice of neighbourhoods involved in the patching if
$\Sigma$ has an even boundary, and so, in particular if it is closed and
 hence a cycle.
Comparison of the two versions of the quantum Stokes' relation 
 makes clear  that the path dependent phase factor [9],
 such as occurs on the right hand
side, is not an autonomous object but is tied to a particular
 sort of wave function.

 If $\Sigma$ is a cycle,
 the path dependent phase factor on the right hand side
collapses to unity leaving the modified flux quantisation condition
$$q\int_{\Sigma}F\quad \in\quad 2\pi\hbar(\z+{w(\Sigma)\over2}).$$
So,  if $w(\Sigma)$ is an odd integer, fluxes are half integer and,
 in particular, cannot vanish.

 If the charge
of the electrically charged spin $1/2$ particle is taken to vanish, the
quantum Stokes' relation reduces to the identity $1=(-1)^{w(\Sigma)}$.
Thus the limit cannot be taken when $w(\Sigma)$ is odd and 
this means that charge neutral
spinor wave functions are forbidden on $\m$. Mathematicians recognise
this as the Stiefel-Whitney obstruction. On four-dimensional
space-times it is unnecessary to use this theory
because  something special happens there, namely that
$w(\Sigma)$ has a simple  geometrical interpretation as being
equal to the self-intersection number  of $\Sigma$, modulo $2$.
 This will be shown 
to follow from something physicists would readily accept because of its
relation to the theory of chiral anomalies, namely the Atiyah-Singer
 index theorem.
\bigskip
\ni{\bf Intersection Numbers and Intersection Matrices}
\bigskip
Two cycles of complementary dimension, 
that is summing to the dimension of the background space-time, 
 generically intersect at
a finite number of discrete points. Using the background orientation,
 these 
points can be assigned a sign. The algebraic
sum of these signs over the points of intersection
 defines the intersection number in  a way that it is unaffected
by homology.
It is instructive to visualise this in the case of the two-torus when a pair of one-cycles are complementary. As $b_1=2$ there are two
 natural one-cycles to consider and these intersect at one point. To define
a self-intersection number two copies of the same one-cycle are considered.
Of course these intersect at all their points but this is 
remedied by displacing
one copy to a homologous cycle and considering points of intersection of
this with the other copy. This can always be done in such a way that 
there are no intersections. In fact the self-intersection number of
  any one-cycle
on a two-manifold always vanishes. On a four-manifold a pair of
 two-cycles are complementary and so self intersection numbers
 can be defined, and in this case do not necessarily vanish.

As explained above, the homology classes of the
 physically relevant two-cycles on a four-manifold
space-time form a lattice. Choosing a basis for it,
 $\Sigma_1,\Sigma_2,\dots \Sigma_{b_2}$ we can define a
$b_2\times b_2$ matrix $Q^{-1}$ formed of the intersection numbers
$$\left(Q^{-1}\right)_{ij}=I(\Sigma_i,\Sigma_j).$$
It is yet another consequence of Poincar\'e duality that this matrix is
unimodular, that is, has determinant equal to $\pm1$. So both $Q^{-1}$ and
its inverse $Q$ have integer entries. In addition $Q$ is symmetric
(whereas the $b_1 \times b_1$ analogue for two-manifolds is antisymmetric, 
thereby explaining why all self-intersection numbers vanish there). So
the lattice of two-cycle homology classes on $\m$ is furnished
 with a scalar product.

Furthermore, $Q$ must be diagonalisable, being real and symmetric,
 and $b_2^+$ and
$b_2^-$ can be defined as
the number of its  positive and negative eigenvalues respectively. 
This  completes the definition of the Hirzebruch signature as 
$b_2^+-b_2^-$, 
mentioned above.

Such unimodular matrices fall naturally into two classes, called even or odd,
according as their diagonal entries are all even or not. Examples with
$b_2^-=0$ are, respectively, the Cartan matrix for the $E_8$ Lie algebra
 and the unit matrix. In fact, when neither of $b_2^{\pm}$ vanish, all
odd  integer unimodular matrices are,
after a change of basis, given by a diagonal matrix with $b_2^+$ entries $1$ and $b_2^-$
entries $-1$ on the diagonal. Even unimodular matrices
only occur when
 the signature $b_2^+-b_2^-$ is  a multiple of eight
and can be constructed from the odd unimodular matrix there
by a construction presented in the penultimate section below.
When either of $b_2^{\pm}$ vanishes
 no corresponding classification theorem is known.

Given  a Maxwell field strength as a closed two-form $F$, it is natural
to form the exterior product $F\wedge F$ which provides a closed four-form
which can be integrated over $\m$. For non-abelian gauge theories
the result is familiar as the instanton number (once the necessary trace is taken). Maxwell theory has no instanton number and instead the result is quadratic in the magnetic fluxes:
$$\int_{\m}F\wedge F=\sum_{ij}\int_{\Sigma_i}F\,\,\,Q_{ij}\,\int_{\Sigma_j}F. $$
The analogue of this for two-manifolds is known as the Riemann 
bilinear identity, and fundamental in the theory of Riemann surfaces.
Further simplification follows on insertion of the quantised values of
the fluxes according to whichever of the two conditions above,
 scalar or spinor,  is appropriate.
Notice how a nontrivial value requires $b_2$ to be nonzero,
 that is a space-time
with non-trivial topological structure with which
 to capture the magnetic fluxes.

\bigskip
\ni {\bf Relation between $w(\Sigma)$ and $Q$ from the Index Theorem}
\bigskip
Four-dimensional space-times of the type considered
 possess yet another special property not valid
in higher dimensions.
 Although it is not always possible to support charge neutral
spinor wave functions it is always possible
 to support charged spinor wave functions
providing the background electromagnetic
 field satisfies the flux quantisation conditions
above. Mathematicians say that there always exists a spin$_C$ structure
but not necessarily a spin structure.

Because of this it is always possible to formulate a Dirac operator $D_A$
including a minimal coupling to $A$, and
 try to solve the Dirac equation $D_A\psi=0$.
The index of $D_A$ is the difference between the numbers of solutions
of opposite chirality, and according to the Atiyah-Singer index theorem,
given by
$$\hbox{INDEX}(D_A)={1\over2}\left({q\over2\pi\hbar}\right)^2\int_{\m}F\wedge F-{\eta(\m)\over8}.$$
We have already met both terms on the right hand side: the second
 involves the Hirzebruch signature and the first the integral just calculated. Inserting the spinor version of the Dirac quantisation conditions yields
$$={1\over2}\sum_{ij}(m_i+{w_i\over2})Q_{ij}(m_j+{w_j\over2})-{\eta\over8},$$
where $w_i=w(\Sigma_i)$ and the integer $m_i$ is
 determined by the flux through $\Sigma_i$. In view of what has been said
this index expression has to be an integer for all values of the integers $m_i$
parametrising the consistent background fields, despite the
 non-integral nature
of the two individual terms. By a trivial rearrangement
 and an abbreviation of notation
$$={1\over2}(mQm+mQw)+{wQw-\eta\over8}\in\z.$$
Thus the integrality of the index  reduces to the two conditions
$$wQw-\eta\in8\z,$$
$$mQm+mQw\in 2\z, \quad\hbox{for all}\quad m_i\in\z.$$
In the theory of integer  unimodular matrices the second condition
is recognised as the definition of $w$ being  what is called
 the characteristic vector of $Q$. There is harmless ambiguity
 in this definition
that amounts to an even integer in each component. This means that $wQw$ is
unambiguous modulo $8$, in accordance with the first condition.
 Of course if $Q$ is even, the zero vector
is a legitimate characteristic vector and the first condition reduces to 
a previous statement, that the signature of $Q$ is a multiple of eight
if it is an even unimodular matrix.

It is a trivial piece of algebra to verify that the solution of the second equation is $w_i=(Q^{-1})_{ii}$, at least modulo two,
 which is precisely what we want. Thus the components of the characteristic
vector of $Q$ are given by the diagonal elements of its inverse, namely
the self-intersection numbers of the $\Sigma_i$. 
More generally, for any cycle $\Sigma$,
$$w(\Sigma)=I(\Sigma,\Sigma)+2\z.$$
that is the self-intersection number as claimed earlier.
More generally, the  conclusion
is that the matrix $Q$, or equivalently $Q^{-1}$, 
carries the essential topological information.

This sort of argument was introduced by S Hawking and C Pope [10]
when they considered the four-manifold $CP(2)$ which has
$b_2=1$ and hence $Q=\pm1$.
\bigskip
\ni{\bf Maxwell Partition Functions}
\bigskip
The way is ready for tests of quantum electromagnetic duality.
It is familiar that the energy-momentum tensor and the equations
 of motion of Maxwell theory respect an $SL(2,I\!\!R)$ group
of symmetry transformations, even if the Lagrangian does not. 
Since this statement involves mixing $F$ and its Hodge dual, $*F$,
 it is required  that $\m$  be endowed with
 a metric that is Minkowski in nature so that there is a single time,
rather than a Euclidean metric. To find the quantum version it seems reasonable
 to consider the partition function $Z=Tr(e^{-E})$ as  $E$, the energy,
 is invariant classically. Indeed the first intimation of quantum duality,
 in this case  a $Z_2$ version, was found by H Kramers and
 G Wannier [11] by studying the partition function of the Ising model.

According to old ideas of R Feynman [12] and M Kac   the partition
 function $Z$ can be expressed as a Feynman path integral integrated over 
the field degrees of freedom. But in this representation time
 is automatically imaginary because of the form of the exponential
 $e^{-E}$, and furthermore, possesses
periodic boundary conditions because of the trace. So, 
now the partition function reads
$$Z(\tau)=Tr(e^{-E})=\int\dots\int\delta A\,
 e^{{i\over\hbar}W_{EUCLIDEAN}(S_1\times {\cal M}_3)},$$
where the Euclidean action $W_{EUCLIDEAN}$ is integrated over the four-manifold
$S_1\times{\cal M}_3$ with periodic time on the circle,
 all equipped with a Euclidean metric. This is what may
 be called the strict partition function. It is a popular procedure
to consider a more general object, no longer necessarily real and positive,
that may be called the extended partition function.
 In this  the Euclidean action
is obtained by integration over any (Poincar\'e dual) four-manifold,
 $\m$, equipped
with a Euclidean metric,
$$\int\dots\int\delta A\, e^{{i\over\hbar}W_{EUCLIDEAN}(\m)}.$$
The exponent, $i/\hbar$ times the Euclidean  action,
 $W_{EUCLIDEAN}$, is actually
a complex number. The real part is always negative, thereby guaranteeing
a good convergence of the integration, whilst the imaginary part is 
proportional to precisely the integral that has already been evaluated
to yield a quadratic form in magnetic fluxes.

To see this in detail it is preferable to work in terms of a field
strength rescaled
so as to have dimensionless fluxes; $f=qF/\hbar$.
In terms of this
$${1\over\hbar}W_{EUCLIDEAN}(\m)={1\over4\pi}\int_{\m}\,f\wedge\hat\tau f,$$
where $\hat\tau=\tau_1+i\tau_2*={\theta\over2\pi}+{2i\pi\hbar\over q^2}*$
and $*$ is again the Hodge dual. Since  the metric involved
 is now Euclidean $*$ has square plus one. Taking the eigenvalue
$+1$
yields the complex variable
$$\tau=\tau_1+i\tau_2={\theta\over2\pi}+{2i\pi\hbar\over q^2}$$
which encodes the dimensionless couplings which parametrise the theory.
The imaginary part $\tau_2$ is the inverse of the fine structure constant
and $\theta$ the vacuum angle. The dependence of the partition functions
 as functions of this variable can be made remarkably explicit
by an argument based on the semi-classical approximation.
\bigskip
\ni{\bf Semi-Classical Evaluation}
\bigskip
Since the partition functions are expressed as integrals of phase
 factors, they can be evaluated by semi-classical methods as a sum
 of contributions from points of stationary phase.
 As the integral is Gaussian the results can be expected to
be exact as E Verlinde [5] and E Witten [4] first pointed out.

The stationary points of the action are given by 
solutions to Maxwell's equations, $df=0=d*f$, where the rescaled
field strengths are constrained to possess fluxes quantised 
in accordance with the conditions appropriate to the
 support of scalar or spinor wave functions.

Since the metric in the integral is Euclidean, Hodge's
 theorem applies and states that the number of linearly independent
solutions (in the sense of real coefficients) 
equals the second Betti number $b_2$. The following normalisation
determines a basis $f^1,f^2,\dots f^{b_2}$:
$$\int_{\Sigma_j}f^i={q\over\hbar}\int_{\Sigma_j}F^i=2\pi\,\delta_j^i.$$
Solutions respecting the two flux quantisation conditions are, respectively,
$$f=\sum_i\,m_if^i\qquad\hbox{or}\qquad \sum_i\,(m_i+{w_i\over2})f^i,\quad m_i\in\z.$$
The values of the $f\wedge f$ term in the action action can be found
immediately, by inserting the bilinear identity above.
To evaluate the other term it is necessary to realise that $*f^i$
also satisfies Maxwell's equations and hence must be expressible
 in terms of the basis of solutions as
$$*f^i=(GQ^{-1})_{ij}f^j\qquad\hbox{where}\qquad (GQ^{-1})^2=1$$
as the Hodge $*$ squares to unity. $G$ is a symmetric, positive definite
matrix depending on the conformal class of the Euclidean metric.

 Then the contribution to $iW_{EUCLIDEAN}/\hbar$
labelled by the integers $m_i$ is either $i\pi m^T\Omega(\tau)m$
or $i\pi (m+{w\over2})^T\Omega(\tau)(m+{w\over2})$, where
$\Omega(\tau)=\tau_1\,Q+i\tau_2\,G$.
Putting everything together
the two possible partition functions associated with scalar and spinor wave functions are respectively (apart from a constant factor),
$$Z_0(\tau)=\tau_2^{b_1-1/2}\sum_{m_i\in\z}\,e^{i\pi m^T\Omega(\tau)m},$$
$$Z_{{w\over2}}(\tau)=\tau_2^{b_1-1/2}\sum_{m_i\in\z+w_i/2}\,e^{i\pi m^T\Omega(\tau)m}.$$
The prefactor is the contribution of Gaussian fluctuations and 
is the same
for all stationary points [4].
It is these expressions that are sufficiently explicit that the response
to modular transformations can be found.
\bigskip
\ni{\bf Modular Group Action}
\bigskip
 It is immediate that, when $Q$ is even, $Z_0(\tau)$ is
invariant under the effect of
 $T:\tau\rightarrow\tau+1$ whereas, if $Q$ is odd, it is not but instead is
 invariant under
 the effect of $T^2$.
The effect of $S:\tau\rightarrow-1/\tau$ on $Z_0$ can be calculated
 using the Poisson summation formula and exploiting the unimodular
 nature of $Q$ that stems from Poincar\'e duality:
$$Z_0(-1/\tau)=
e^{-{2i\pi\eta\over8}}\tau^{{\chi+\eta\over4}}
\overline{\tau}^{{\chi-\eta\over4}}Z_0(\tau).$$
Thus when $Q$ is even, both $S$ and $T$ act nicely on $Z_0$.
 Since the whole modular group consisting of fractional linear transformations,
$$\tau\rightarrow {a\tau+b\over c\tau+d},\qquad a,b,c,d\in\z,\quad ad-bc=1,$$
is generated by $S$ and $T$, $Z_0$ is then covariant under this action.
 This is like the original $SL(2,I\!\!R)$ action in the classical theory,
 mentioned above,
broken to a discrete subgroup by quantum effects. However when $Q$ is odd
the nice action on $Z_0$ is generated by $S$ and $T^2$, which yields a subgroup
of index three in the full modular group. This is how the matter was left
by E Verlinde and E Witten.

With the extra work above concerning fractional fluxes when $Q$ is odd, the remedy is apparent.
When $Q$ is even, $w$ vanishes and the two partition functions are the same.
When $Q$ is odd they differ and both come into play. A rather general
argument is given
in the next section to show that when both are
 taken into account together with a third, ($Z_0(\tau+1)$), 
the whole modular group action is again realised.

Notice that the $S$ action shows that the action of the modular group is
accompanied by factors with \lq\lq modular weights" ${\chi\pm\eta\over4}$, 
where $\chi$ and $\eta$ are the Euler number and Hirzebruch signature
of the four-manifold $\m$ entering the extended partition function.
These are precisely those topological numbers, 
independent of any choice of metric, which are local. 

All this applies to  the two extended partition functions. 
The strict partition function, $Tr(e^{-E})$, is given as 
the special case when $\m$ factorises
as $S_1\times{\cal M}_3$. Then  both of the numbers, $\chi$ and $\eta$,
  vanish
and $Q$ is even so $Z_0$ and $Z_{{w\over2}}$ coincide.
Furthermore this strict partition function,
 which is the physically meaningful
quantity in Minkowski space, is indeed invariant
under the full modular group.

\bigskip

\ni{\bf Theta Functions and Even Integral Lattices}
\bigskip
In order to understand how the action of the 
modular group relates the two versions of the
partition function which are relevant when $Q$ is odd,
some ideas are developed that seem to have some intrinsic interest.

Suppose a slightly more abstract language is introduced and a
 lattice $\Lambda$ is endowed with a non-singular scalar product.
Then the reciprocal lattice $\Lambda^*$ can be defined
 (so that $\Lambda\cdot\Lambda^*\in\z$). $\Lambda$ is said to be integral
if $\Lambda\subset\Lambda^*$. In this case $\Lambda^*$ decomposes into
cosets with respect to $\Lambda$. These cosets form a finite abelian group
denoted $\Lambda^*/\Lambda\equiv Z(\Lambda)$, say. If $Z(\Lambda)$
possesses only one element then $\Lambda=\Lambda^*$ and is
unimodular.

If $\Lambda$ is even, that is $\ell^2$ is an even integer for any element
$\ell$ of $\Lambda$, and the scalar product is positive definite, then there
is known to be a  nice construction
 for $|Z(\Lambda)|$ theta functions, one
for each coset of $\Lambda$ in $\Lambda^*$, denoted $\lambda_j+\Lambda$ where $\lambda_j$ is a representative element, by
$$\theta_j(\tau)=\sum_{\ell\in\lambda_j+\Lambda}\,e^{\pi\,i\tau\ell^2}.$$
These are holomorphic in the upper half plane of the complex variable $\tau$
and support an action of the modular group acting on $\tau$ by the usual
fractional linear transformations. For example,
the effect of $T:\tau\rightarrow\tau+1$ is simply
$$\theta_j(\tau+1)=e^{\pi\,i\ell^2}\theta_j(\tau).$$
This would not work if the lattice $\Lambda$ were odd.
 But whether it is odd or even
it is not difficult to use the Poisson summation formula to find
 that $S:\tau\rightarrow-1/\tau$ has nice action on these theta functions.
This is sufficient as $S$ and $T$ generate the whole modular group.
These results are well known and explained in the book by Green,
 Schwarz and Witten [13], for example.

The magnetic charge lattice defined previously with scalar product coming from
the intersection matrix, $Q$, differs in that it is actually unimodular,
and hence integral, but more importantly in that the scalar
product need not be positive definite. Indeed $\eta$ measures the signature.
That means that the corresponding theta functions have to be more complicated, exactly as was
found above. Nevertheless the same  sort
of coset construction can again be performed, now
 for even integral indefinite lattices.
The corresponding 
theta functions are no longer holomorphic in the upper half $\tau$ plane
but they are still highly convergent there
 (because of the structure coming from $G$). The action of the modular
group can be evaluated in detail and is very similar to the case of
 positive definite $Q$.

 The $S$ and $T$ actions are not totally independent
because of the relation within the modular group, $(ST)^3=-1$. 
The consistency of this leads to an identity, known as Milgram's formula
in the context of the theory of even lattices [14],
$$\sum_{j=1}^{|Z(\Lambda)|}\,e^{i\pi\lambda_j^2}=\sqrt{|Z(\Lambda)|}e^{{2i\pi\eta\over8}}.$$

Now return to the situation that $Q$ is an odd unimodular matrix.
Corresponding to it is an odd unimodular lattice, $\Lambda$ and two
partition functions, $Z_0$ and $Z_{w/2}$, relevant
 to the fluxes supporting complex scalar and spinor
 wave functions respectively. They involve sums
 over $\Lambda$ and $\Lambda+{w\over2}$ in the new notation. Consider the union
of these two sets:
$$\Lambda_{TOTAL}=\l\cup(\l+{w\over2}).$$
This is closed under addition as $\l$ is and $w$, being its
 characteristic vector, lies in it. This means $\Lambda_{TOTAL}$ is a lattice.
$\l$ itself can be split into two pieces,
 $\l_{EVEN}$ and $\l_{ODD}$ according as the
 scalar product with $w$ is even or odd.
 Then $\l_{EVEN}$ is an even lattice and is reciprocal to $\l_{TOTAL}$
$$(\l_{EVEN})^*=\l_{TOTAL}.$$
So we have an example of the situation considered above with an even lattice
whose reciprocal could be decomposed into
 $|Z(\l_{EVEN})|=4$ cosets, each with their own theta function.
Here
$$\l_{TOTAL}=\l_{EVEN}\cup\l_{ODD}\cup(\l_{EVEN}+{w\over2})\cup(\l_{ODD}+{w\over2}).$$
Corresponding to this are four theta 
functions spanning the space upon which the full modular group acts.
 Two linear combinations yield the two partition
 functions $Z_0$ and $Z_{{w\over2}}$ and that 
is why the modular group acts on them.
 Actually a certain linear combination of the four  theta functions vanishes,
 leaving just three needed to support the modular group action,
 as claimed earlier.

Milgram's formula above can be specialised to $\l_{EVEN}$ and implies
 that $w^2-\eta$ should vanish, mod $8$, the result that was previously deduced
from the index theorem.
\bigskip
\ni{\bf More on Odd Unimodular Lattices}
\bigskip
Although not strictly relevant to the main argument it
 is interesting to develop
the above construction of an even integral lattice $\l_{ODD}$ from an odd unimodular
lattice $\l$ a step further.
Consider two other subsets of the above coset decomposition of $\l_{TOTAL}$,
$$\l'\equiv\l_{EVEN}\cup(\l_{EVEN}+{w\over2})\quad\hbox{and}
\quad\l''\equiv\l_{EVEN}\cup(\l_{ODD}+{w\over2}).$$
By Milgram's formula, when the signature, $\eta$, is even,
 the characteristic vector $w$ lies in $\l_{EVEN}$, 
 and as a consequence,
 $\l'$ and $\l''$ are both lattices.
Furthermore they are integral, unimodular  lattices if $\eta$ is a multiple of
$4$. These are even when $\eta$ is a multiple of $8$ and odd otherwise.
This procedure is interesting because it can generate non-trivial
unimodular lattices from trivial ones, for example, the $E_8$ root lattice
from the hypercubic lattice $\z^8$.

Corresponding to this geometry the modular behaviour of the theta functions
simplifies as explained in [3].
\bigskip
\ni{\bf Discussion}
\bigskip
The arguments presented to support the physical 
ideas of quantum electromagnetic duality
 have illustrated what seems to be a beautiful 
interplay between  different mathemematical ideas: homology theory,
 spin obstructions, index theorems,
theory of integral lattices, 
theta functions and the modular group,  and so on.
This is reassuring but it indicates that there is a deeper
underlying structure still to be found. 

Unfortunately the test of quantum electromagnetic duality 
considered is  rather crude
and it
is therefore desirable to formulate more stringent tests. It is odd that the most interesting
calculations involve what were called the extended partition functions. Except for 
the strict partition function which occurred as a special case, these have a rather unclear
physical interpretation.  For example, what is the physical 
significance of the Euclidean metric used in the construction of the Euclidean action?
It would be wrong to think of it as being obtained from a Minkowski metric 
by some sort of Wick rotation. Indeed the Euler number $\chi(\m)$
need not vanish, and this would leave no
possibility for a Minkowski metric on $\m$.

The work has relied on special features of four-manifolds not valid in higher
dimensions, but  these have been equipped with 
Maxwell fields which, being two-forms,  are mid-forms. It can be anticipated
 that something similar happens in space-times of higher even dimension
when, again,  mid-forms are considered together with their putative 
 coupling to the appropriate branes, instead of particles.
 But this needs checking and, indeed,  requires  new mathematics
as a good way of considering brane wave functions seems to be lacking so far.

In particular, the success of the work so far has depended on careful attention
to signs associated with spinor structures and it is important to understand
how the simple geometric interpretation valid on four-manifolds 
can be extended to higher dimensions.
 It is even more difficult to see how to extend
the analysis to superstring theories with the same 
level of precision.

\bigskip
I am grateful to Miguel Virasoro for organising such a pleasant meeting.
I wish to thank Marcos Alvarez for his collaboration in  respect of the work
in [2] and [3]. I also wish to repeat my thanks to those cited there. Finally
my thanks for their helpful comments on the present manuscript go to Stephen Howes,
Luis Miramontes and Ani Sinkovics.
\bigskip
\def\np{Nucl.~Phys.}

\def\pr{Phys.~Rev.}
\def\cmp{Commun.~Math.~Phys.}
\def\prs{Proc.~Roy.~Soc.}

\medskip
\ni [1] DI Olive; \lq\lq Exact Electromagnetic Duality" \np (Proc Suppl)
{\bf 45A} (1996) 88-102.
\smallskip
\ni [2] M Alvarez and DI Olive; \lq \lq The Dirac quantisation
condition for fluxes on Four-manifolds", 
\cmp {\bf 210} 
(2000) 13-28, {\tt hep-th/9906093}.
\smallskip
\ni [3] M Alvarez and DI Olive; \lq \lq 
Spin and Abelian Electromagnetic Duality on Four-manifolds", \cmp {\bf 217} 
(2001) 331-356, {\tt hep-th/0003155}.
\smallskip
\ni [4] E Witten; \lq\lq On S-duality in abelian gauge theory",
 Selecta Math (NS) {\bf 1} (1995), 383-410, {\tt hep-th/9505186}.
\smallskip

\ni [5] E Verlinde; \lq\lq Global aspects of electric-magnetic duality",
\np\ {\bf B455} (1995), 211-228.
\smallskip

\ni [6] O Alvarez; \lq\lq Topological Quantisation and Cohomology",
\cmp\ {\bf 100} (1985) 279-309. 
\smallskip

\ni [7] PAM Dirac; \lq\lq Quantised singularities in the
electromagnetic field", \prs\ {\bf A133} (1931) 60-72.

\smallskip
\ni [8] TT Wu and CN Yang; \lq\lq Concept of non-integrable phase factors
and global formulation of gauge fields", \pr\ {\bf D12}
(1975), 3845-3857.
\smallskip
\ni [9] PAM Dirac;
\lq\lq Gauge invariant formulation of Quantum Electrodynamics'',  Canadian
Journal of Physics {\bf{33}} (1955), 650-660.
\smallskip
\ni [10] SW Hawking and CN Pope; \lq\lq Generalised spinor structures 
in quantum gravity",  Phys Lett {\bf B73} (1978) 42-44.

\ni [11] HA Kramers and GH Wannier; \lq\lq Statistics of the Two-Dimensional
Ferromagnet I",  \pr {\bf 60} (1941), 252-276.
\smallskip

\ni [12] RP Feynman and AR Hibbs; {\sl Quantum Mechanics and Path
Integrals}, Chapter 10,  McGraw-Hill, 1965,
\smallskip
\ni RP Feynman; {\sl Statistical Mechanics}, Chapter 3,
Benjamin/Cummings, 1972.
\smallskip
\smallskip
\ni [13] M Green, J Schwarz and E Witten; {\sl Superstring Theory},
Vol.~ 2, Appendix 9B, Cambridge University Press, 1987.
\smallskip
\smallskip
\ni [14] J Milnor, D.~Husemoller; {\sl Symmetric bilinear 
forms}, Ergebnisse der Mathematik und ihrer Grenzgebiete, Band 73, 
Springer-Verlag, 1973.

\bye